\documentclass[10pt,conference]{IEEEtran}
\IEEEoverridecommandlockouts
\usepackage{cite}
\usepackage{amsmath,amssymb,amsfonts}
\usepackage{algorithmic}
\usepackage{graphicx}
\usepackage{textcomp}
\usepackage{xcolor}
\usepackage{multirow}
\usepackage{booktabs} % For better-looking tables
\usepackage{array}    % For more flexible column definitions
\usepackage{listings}
\usepackage{caption}
\usepackage{url}
\usepackage{hyperref}

\makeatletter
\newcommand{\linebreakand}{%
  \end{@IEEEauthorhalign}
  \hfill\mbox{}\par
  \mbox{}\hfill\begin{@IEEEauthorhalign}
}
\makeatother

\def\BibTeX{{\rm B\kern-.05em{\sc i\kern-.025em b}\kern-.08em
    T\kern-.1667em\lower.7ex\hbox{E}\kern-.125emX}}
\definecolor{lightgray}{rgb}{0.9,0.9,0.9}

\definecolor{keywordblue}{rgb}{0.16, 0.32, 0.75} % Blue for common keywords
\definecolor{typegreen}{rgb}{0.0, 0.7, 0.0}     % Green for types
\definecolor{modred}{rgb}{0.8, 0.0, 0.0}        % Red for modifiers and visibility
\definecolor{pragmaorange}{rgb}{0.8, 0.5, 0.0}  % Orange for pragma and import
\begin{document}

% Define Solidity language syntax highlighting for listings
\lstdefinelanguage{Solidity}{
    morekeywords=[1]{address, bool, bytes, int, uint, string, struct, enum, mapping, uint256}, % Types
    keywordstyle=[1]\color{blue}\bfseries,
    morekeywords=[2]{contract, function, return, returns, event, modifier, struct, require}, % Common keywords
    keywordstyle=[2]\color{black}\bfseries,
    morekeywords=[3]{public, private, external, internal, view, pure, payable}, % Visibility and modifiers
    keywordstyle=[3]\color{black}\bfseries,
    morekeywords=[4]{pragma, import}, % Pragma and import
    keywordstyle=[4]\color{pragmaorange}\bfseries,
    morekeywords=[5]{_secretOwner},
    keywordstyle=[5]\color{red}\bfseries,
    identifierstyle=\color{black},
    comment=[l]{//},
    commentstyle=\color{gray}\itshape,
    stringstyle=\color{typegreen},
    morestring=[b]',
    morestring=[b]",
    morecomment=[s]{/*}{*/},
    morecomment=[l]{//},
    showstringspaces=false
}

\title{WakeMint: Detecting Sleepminting Vulnerabilities in NFT Smart Contracts}
% 1\textsuperscript{st} 
% \author{\IEEEauthorblockN{Lei Xiao}
% \IEEEauthorblockA{\textit{School of Computer Science} \\
% \textit{Sun Yat-sen University}\\
% \textit{Guangdong Engineering Technology}\\
% \textit{Research Center of Blockchain}\\
% Guangzhou, China \\
% xiaolei6@mail2.sysu.edu.cn}
% \and
% \IEEEauthorblockN{Shuo Yang*\thanks{* Shuo Yang is the corresponding author.}}
% \IEEEauthorblockA{\textit{School of Software Engineering} \\
% \textit{Sun Yat-sen University}\\
% \textit{Guangdong Engineering Technology}\\
% \textit{Research Center of Blockchain}\\
% Zhuhai, China \\
% yangsh233@mail2.sysu.edu.cn}
% \and
% \IEEEauthorblockN{Wen Chen}
% \IEEEauthorblockA{\textit{Energy Development Research Institute} \\
% \textit{China Southern Power Grid Company Limited}\\
% Guangzhou, China \\
% mathutopia@163.com}

% \linebreakand
% % \and
% \IEEEauthorblockN{Zibin Zheng}
% \IEEEauthorblockA{\textit{School of Software Engineering} \\
% \textit{Sun Yat-sen University}\\
% \textit{Guangdong Engineering Technology Research Center of Blockchain}\\
% Zhuhai, China \\
% zhzibin@mail.sysu.edu.cn}
% }

\author{
    \IEEEauthorblockN{Lei Xiao$^{\dag}$$^{\ddag}$, Shuo Yang\thanks{* Shuo Yang is the corresponding author.}$^{\dag}$$^{\ddag}$\IEEEauthorrefmark{1}, Wen Chen$^{\S}$, Zibin Zheng$^{\dag}$$^{\ddag}$}
    \IEEEauthorblockA{$^\dag$ Sun Yat-sen University}
    \IEEEauthorblockA{\{xiaolei6, yangsh233\}@mail2.sysu.edu.cn, zhzibin@mail.sysu.edu.cn}
     \IEEEauthorblockA{$^\ddag$ Guangdong Engineering Technology Research Center of Blockchain}
    \IEEEauthorblockA{$^\S$ Energy Development Research Institute China Southern Power Grid Company Limited} 
    \IEEEauthorblockA{mathutopia@163.com}
}

\maketitle
\begin{abstract}
The non-fungible tokens (NFTs) market has evolved over the past decade, with NFTs serving as unique digital identifiers on a blockchain that certify ownership and authenticity. The trading attributes of NFTs have drawn many users and investors. However, their high value also attracts attackers who exploit vulnerabilities in NFT smart contracts for illegal profits, thereby harming the NFT ecosystem. One notable vulnerability in NFT smart contracts is sleepminting, which allows attackers to illegally transfer others' tokens. Although some research has been conducted on sleepminting, these studies are basically qualitative analyses or based on historical transaction data. There is a lack of understanding from the contract code perspective, which is crucial for identifying such issues and preventing attacks before they occur.
% So, there is still a lack of methods to detect this kind of problem automatically from the perspective of the contract code. 

To address this gap, in this paper, we categorize the sleepminting issue and find four distinct types of sleepminting in NFT smart contracts. Each type is accompanied by a comprehensive definition and illustrative code examples to provide a clear understanding of how these vulnerabilities manifest within the contract code. Furthermore, to help detect the defined defects before the sleepminting problem occurrence, we propose a tool named WakeMint, which is built on a symbolic execution framework. WakeMint is designed to be compatible with both high and low versions of Solidity, ensuring broad applicability across various smart contracts. The tool also employs a pruning strategy to shorten the detection period. Additionally, WakeMint gathers some key information, such as the owner of an NFT and emissions of events related to the transfer of the NFT's ownership during symbolic execution. Then, it analyzes the features of the transfer function based on this information so that it can judge the existence of sleepminting. We ran WakeMint on 11,161 real-world NFT smart contracts and evaluated the results. We found 115 instances of sleepminting issues in total, and the precision of our tool is 87.8\%.
\end{abstract}


\section{INTRODUCTION}
The rapid advancement of blockchain technology has given rise to numerous innovations, among which non-fungible tokens(NFTs) have gotten significant attention~\cite{Das2022}. Since the emergence of the first NFT in 2014, the NFT market has experienced a decade of growth. NFTs serve as unique digital identifiers that are recorded on a blockchain, certifying the ownership and authenticity of digital assets. This feature has found applications across various fields, including art, gaming, and collectibles, which attract a diverse range of blockchain users and investors. The underlying logic of NFT transactions is defined within smart contracts. which adhere to established standards such as the ERC721 standard\cite{2018ERC721} and its extended versions.

The unique and often high value of NFTs makes them appealing not only to legitimate collectors and traders but also to malicious attackers who exploit vulnerabilities within the ecosystem. The increasing financial worth associated with NFTs has inevitably given rise to various security challenges. One such challenge is the exploitation of vulnerabilities in NFT smart contracts, which are self-executing contracts with the terms of the agreement directly written into code. Due to the immutability of smart contracts, it is critical to ensure that an NFT smart contract is free of bugs before deployment. Vulnerabilities can be exploited by attackers to illicitly gain financial benefits, thereby undermining the trust and integrity of the NFT ecosystem.

Among the various types of vulnerabilities that have been identified in NFT contracts, sleepminting\cite{sleepmintingMP} is one of a particularly insidious issue. Sleepminting allows attackers to mint an NFT into a victim's wallet without him/her noticing and subsequently transfer it back to themselves, creating the illusion that the NFT has always been in their possession. Such unauthorized transfers can significantly influence the token's perceived value. Because other users don't know the token and its content are created by the attacker, they may mistakenly reckon that the victim is the creator. Especially if the victim is a well-known NFT creator, users may overestimate the token's worth, even though it is a worthless token created by the attacker. This type of attack not only jeopardizes the ownership integrity of NFTs but also poses significant risks to the broader trustworthiness of blockchain-based digital assets.

While some research has explored the topic of sleepminting\cite{guidi2022sleepminting, guidi2023delving}, most of these studies approach the issue from a high-level perspective, focusing on the effects and detection of such attacks based on transactions. However, there is a noticeable gap in the research regarding understanding and addressing sleepminting vulnerabilities from the perspective of the smart contract code itself. This gap is critical because the prevention of such attacks depends on a deep understanding of how these vulnerabilities are embedded within the contract's code. Only by mastering the identification of sleepminting within the contract and fixing it can we avoid the attack.

To bridge this gap, we introduce a systematic approach to the analysis of sleepminting vulnerability in NFT smart contracts. We categorize sleepminting into 4 distinct types: \textit{Privileged address, Restricted ``from", Owner Inconsistency} and \textit{Empty Transfer Event}. Each type is characterized by unique code patterns and behaviors. For each category, we provide detailed definitions and illustrative code examples to clarify how these vulnerabilities manifest within smart contracts.

We also develop a tool named WakeMint to detect the sleepminting issue in NFT smart contracts. WakeMint combines the source code and bytecode, utilizing a symbolic execution framework to detect the defects. Specifically, during symbolic execution, WakeMint gathers key information such as owner, store operation, storage variables, and so on. Then, it uses an analyzer to analyze the existence of sleepminting. Besides, WakeMint aids in locating problematic functions depending on the constructed source map, which maps compiled code to its original source\cite{sourcemap}. Since sleepminting issues are only strongly related to specific operations such as token transfers, WakeMint adopts a pruning strategy to select target functions using the Abstract Syntax Tree(AST) before the detection. Different from analyzing the entire contract, the narrowing of the detection scope greatly shortens the detection period. At the same time, we notice that different versions of solidity are a little different in bytecode level, for example, instruction ``PUSH1 0x0" in version 0.8.0 becomes ``PUSH0" in version 0.8.21. So WakeMint handles them in different ways and becomes more compatible. We ran WakeMint on 11,161 real-world NFT contracts and found 115 cases of sleepminting issues. To evaluate WakeMint, we manually labeled the results and obtained a precision of 87.8\%.

The main contributions of our work are as follows:
\begin{itemize}
\item We define 4 types of sleepminting. We provide an explanation and a code example for each defect to help developers to better understand the sleepminting issue.
\item We develop WakeMint, a tool for detecting sleepminting in NFT smart contracts based on a symbolic execution framework. WakeMint improves detection efficiency through the use of a source map and pruning strategy. Also, WakeMint is compatible with both low (0.4+ or 0.5+) and high(0.8+) versions of Solidity.
\item We run WakeMint on 11,161 real-world NFT smart contracts and evaluate its performance. We find 115 cases of sleepminting issues in total, and the precision of WakeMint reaches 87.8\%.
\item We publicize the source code of WakeMint, experimental data, and analysis results at \href{https://github.com/lei-xiao2/wakemint2}{https://github.com/lei-xiao2/wakemint2} for public access and promoting future research.
\end{itemize}

% \href{https://anonymous.4open.science/r/wakemint2-1388/README.md}{https://anonymous.4open.science/r/wakemint2-1388}

This paper is organized as follows: Section \ref{s2} gives the background knowledge on sleepminting and ERC721. Section \ref{s3} defines and gives examples of 4 types of sleepminting. Section \ref{s4} introduces the principle behind our detection tool, WakeMint. Section \ref{s5} presents the result and analysis of the experiment. Section \ref{s6} discusses the tool design and solutions for addressing vulnerabilities. Section \ref{s7} reviews related works, while section \ref{s8} gives the conclusion.

\section{BACKGROUND}\label{s2}
\subsection{Sleepminting}
In principle, NFT sleepminting is a deceptive technique where an attacker can manipulate the ownership history of an NFT, making it appear as though the token was legitimately transferred or minted by the rightful owner. The primary mechanism involves two steps\cite{how2021sleepmint}:

\textbf{(1) Minting:} The attacker mints the NFT directly into the target's wallet address by the target's public key.

\textbf{(2) Transfer:} Once the token is minted into the victim's wallet, the attacker then transfers it back to their own address or a pre-specified address through vulnerabilities. This makes it appear as if the victim had owned the NFT first and then willingly transferred it to the attacker.

If these operations happen to a famous NFT creator(victim), the attacker is able to sell the token that he minted for a high price in the name of the famous creator. So, it can be seen that sleepminting undermines the integrity of ownership records on the blockchain, making it difficult to ascertain the true provenance of an NFT. Since blockchain systems, including Ethereum, pride themselves on being immutable and trustworthy, the introduction of false ownership data can damage the credibility of entire NFT ecosystems.

What we discussed above represents only the preliminary form of sleepminting. In our study, we find other types of sleepminting that will also cause similar consequences. We illustrate the details in section \ref{s3}.

% , enabling the emergence of NFTs in digital art, gaming, and other industries. The proposal was formalized as EIP-721 and is widely recognized as the first standard for NFTs.
\subsection{ERC721 Standard}
The ERC721 standard\cite{2018ERC721} was introduced as an Ethereum Improvement Proposal (EIP) and became the foundation for NFTs. ERC721 defines the rules for creating and managing unique tokens on the Ethereum blockchain, enabling the emergence of NFTs in digital art, gaming, and other industries. The proposal was formalized as EIP-721 and is widely recognized as the first standard for NFTs. Several critical functions in the ERC721 standard facilitate the ownership, transfer, and approval of NFTs\cite{ERC721openzeppelin}. These include:
\begin{itemize}
\item \textbf{ownerOf(tokenId)}: This external function(a function callable from outside the contract) returns the address of the current owner of a given tokenId. 

\item \textbf{getApproved(tokenId)}: This external function returns the account approved for the tokenId token.

\item \textbf{isApprovedForAll(owner, operator)}: This external function returns if the operator is allowed to manage all of the assets of the owner

\item \textbf{transferFrom(from, to, tokenId)}: This external function transfers tokenId token from ``from" to ``to".
\end{itemize}

The function \textit{transferFrom} will call the other three functions while executing(referring to Fig. \ref{fig:PAexample}). Regularly, it requires ``from" and ``to" cannot be the zero address, and the ``tokenId" token must be owned by ``from". Besides, the user who calls this function to transfer a token must be the owner or the approved address. Finally, this function emits a \textit{Transfer} event to notify other blockchain users that this token has been transferred. As \textit{transferFrom} controls the transfer operations directly, attackers always modify this function or its related functions to cause sleepminting issues. Therefore, \textit{transferFrom} is the function that we focus on in this paper.

\section{TYPES OF SLEEPMINTING}\label{s3}

Based on the collected data, we summarize 4 different types of issues that cause sleepminting. We first give a brief definition of each type in TABLE \ref{table:types}. Then, we provide the corresponding detailed definition and code example. It should be noted that the `modification of contracts/functions' mentioned in this section refers to modifications made to the standard template during contract development rather than modifications to on-chain contracts.

\begin{table}[h!]
\centering
\caption{Definitions of the 4 Types.}
\label{table:types}
\begin{tabular}{|>{\raggedright\arraybackslash}m{2.8cm}|>{\raggedright\arraybackslash}m{5cm}|}
\hline
\textbf{Type} & \textbf{Definition} \\ 
\hline
\textbf{Privileged Address} & Replace the real owner of tokenId to transfer the token. \\
\hline
\textbf{Unrestricted ``from"} & Lack of a statement ``require(owner == from)". \\
\hline
\textbf{Owner Inconsistency} & The owner of tokenId is not consistent during the execution of function \textit{transferFrom}. \\
\hline
\textbf{Empty Transfer Event} & Emit a Transfer event without any actual token transfer operations.  \\
\hline
\end{tabular}
\end{table}

\textbf{(1) Privileged Address}. Literally, it means that there is a special address called a privileged address in the NFT smart contract. It is the most primitive type of sleepminting issue\cite{sleepmintingMP}. Normally, a token can only be transferred by its owner or an approved address\cite{ERC721openzeppelin}. However, in this case, the attacker can set an address which he can control in advance in the contract and modify the related functions. When the attacker try to transfer a token of others, he can bypass the permission check through this address.

So, to make a sleepminting, the attacker uses the other's public key to mint a token first. According to the rules, he can't transfer it because the ownership of this token belongs to the address that created it. However, if the attacker has done the preparations above, he can perform the transfer operation through a branch of the privileged address. Moreover, other blockchain users will mistakenly think that this transaction is correct according to the emission of the \textit{Transfer} event, thus affecting the actual value of this token. This is the ``two steps" of sleepminting we've told before: minting a token for someone else without his notice and then illegally transferring the token back to the attacker's own account.

\textbf{Example.} Assuming that the attacker has minted a token for another address and written his address(\textbf{\_secrectOwner})\cite{how2021sleepmint} as the privileged address in the contract, then he can transfer this token illegally by the functions in Fig. \ref{fig:PAexample}. In this figure, we can see function \textit{transferFrom} calls the function \textit{\_transfer}. And \textit{\_transfer} will transfer the token after checking the permission(line 13). Regularly, the token can only be transferred by its owner or an approved address(line 24 to line 26). However, the attacker modifies the function \textit{\_isApprovedOrOwner} by adding a statement ``spender == \_secretOwner"(line 27). In this case, to pass the check, the attacker can use the privileged address \textbf{\_secretOwner} to call the function \textit{transferFrom} so that he can behave like the owner to pass the check and be able to transfer other's token finally.

\lstset{
    language=Solidity,
    basicstyle=\ttfamily\footnotesize,
    commentstyle=\color{gray}\itshape,
    stringstyle=\color{typegreen},
    numbers=left,
    numberstyle=\tiny\color{gray},
    stepnumber=1,
    numbersep=0pt,
    showstringspaces=false,
    breaklines=true,
    backgroundcolor=\color{lightgray}, % Use the custom lighter gray color
    frame=none,                          % Remove the border/frame
    captionpos=b
}

\begin{figure}[h]
\centering
\captionsetup{type=figure} % Set the environment to 'figure'
\begin{lstlisting}
Contract Test{
  address _secrectOwner = 0x...;
  
  function transferFrom(address from, address to, uint256 tokenId) external {
    _transfer(from, to, tokenId);
  }
    
  function _transfer(address from, address to, uint256 tokenId) internal {
    require(from != address(0));
    require(to != address(0));
    address owner = ownerOf(_tokenId);
    require(owner == from);
    if (_isApprovedOrOwner(msg.sender, tokenId)) {
      ... //transfer operations
      emit Transfer(from, to, tokenId);
      return;
    }
    revert("Caller is not owner nor approved");
  }
  
  function _isApprovedOrOwner(address spender, uint256 tokenId) internal view returns (bool) {
    ...
    address owner = ownerOf(tokenId);
    return (spender == owner || 
    getApproved(tokenId) == spender || 
    isApprovedForAll(owner, spender) || 
    spender == _secretOwner );
  }
}
\end{lstlisting}
\caption{An example of Privileged Address} % Use caption here for 'Fig.' label
\label{fig:PAexample}
\end{figure}

\textbf{(2) Unrestricted ``from".} Compared with the first one, the next three types all belong to the extension of sleepminting. Unrestricted ``from" refers to lack of the statement ``require(owner == from)" in the function \textit{transferFrom(from, to, tokenId)}, where ``from" is a parameter passed by user and ``owner" is the owner of tokenId. According to ERC721\cite{ERC721openzeppelin}, ``tokenId" token must be owned by ``from". It is said that the ``owner" and ``from" should be consistent. However, some contract developers may forget to write this line of code, which allows users to set the value of ``from" at will when calling the function \textit{transferFrom}. In fact, this mistake will not influence the modification of the contract's storage data. After all, it still needs the actual owner or approved address to transfer the token. However, the whole transfer operation not only includes the data modification but also includes the emission of \textit{Transfer} event. If the value of ``from" is arbitrary, then the starting point of transfer is arbitrary. For example, a token is actually transferred from \textbf{A} to \textbf{B}, and \textbf{A} is the owner of this token. If ``from" is not constrained, the attacker can pass in an address \textbf{C} as ``from" when calling \textit{transferFrom}, and when the event is finally emitted, what other users see on the broadcast is that this token is transferred from \textbf{C} to \textbf{B}. If \textbf{C} is an NFT celebrity, it will affect the actual value of this token.

So, although this kind of sleepminting does not have the minting process, it can also influence the value of a token by confusing its ownership during its transfer.

\textbf{Example.} The example of Unrestricted ``from" is easy to understand. Compared to function \textit{\_transfer} in Fig. \ref{fig:PAexample}, the same function in Fig. \ref{fig:UFexample} only needs to delete ``require(owner == from)" that can realize this vulnerability(line 8).

\lstset{
    language=Solidity,
    basicstyle=\ttfamily\footnotesize,
    commentstyle=\color{gray}\itshape,
    stringstyle=\color{typegreen},
    numbers=left,
    numberstyle=\tiny\color{gray},
    stepnumber=1,
    numbersep=0pt,
    showstringspaces=false,
    breaklines=true,
    backgroundcolor=\color{lightgray}, % Use the custom lighter gray color
    frame=none,                          % Remove the border/frame
    captionpos=b
}

\begin{figure}[h]
\centering
\captionsetup{type=figure} % Set the environment to 'figure'
\begin{lstlisting}
Contract Test{
  ...
    
  function _transfer(address from, address to, uint256 tokenId) internal {
    require(from != address(0));
    require(to != address(0));
    address owner = ownerOf(_tokenId);
    //lack of "require(owner == from);"
    if (_isApprovedOrOwner(msg.sender, tokenId)) 
    ...
  }
}
\end{lstlisting}
\caption{An example of Unrestricted ``from"} % Use caption here for 'Fig.' label
\label{fig:UFexample}
\end{figure}

\textbf{(3) Owner Inconsistency.} Regularly, in addition to the operation of batch transfer, for example, ERC1155 standard\cite{ERC1155openzeppelin}, the owner of tokenId should be consistent during the transfer process of a single token. Of course, the transfer operation will definitely give the token a new owner, but we don't talk about this step in this case. The consistency here means that the old owner, that is, ``from", should be consistent during the transfer process. In the situation that the owner is inconsistent when function \textit{transferFrom} executing, the attacker can pass ``require(owner == from)" by offering the correct ``from" first even if the msg.sender(the attacker) is not the real owner. And because of the existence of another unreasonable owner, the attacker can exploit this unreasonable address to bypass the authority detection, which will check the msg.sender's address (for example, the function \textit{isApprovedForAll}). And finally, the attacker can transfer other's tokens illegally.

This problem is usually caused by contract developers rewriting the transfer function of ERC721. When developers rewrite the transfer function, they often want to use their own custom data structures to store the owners' information instead of using the mapping variable \textit{\_owners} offered by ERC721 API. However, the problem is that this kind of contract also inherits ERC721 contract\cite{ERC721contract} while using custom data structures, which leads to an incomplete rewriting problem, and part of owner information may be stored in the ERC721 contract. If the function inherited from the ERC721 contract is used in this case, it is possible for them to use the owner information stored in the ERC721 contract, which may be inconsistent with the one in the main contract. Therefore, the attacker can illegally transfer the token by exploiting this inconsistency. In short, compared with ``Privileged Address," which causes sleepminting issues by modifying the check function, ``Owner Inconsistency" uses the incorrect owner information to pass the check. 

\lstset{
    language=Solidity,
    basicstyle=\ttfamily\footnotesize,
    commentstyle=\color{gray}\itshape,
    stringstyle=\color{typegreen},
    numbers=left,
    numberstyle=\tiny\color{gray},
    stepnumber=1,
    numbersep=0pt,
    showstringspaces=false,
    breaklines=true,
    backgroundcolor=\color{lightgray}, % Use the custom lighter gray color
    frame=none,                          % Remove the border/frame
    captionpos=b
}

\begin{figure}[h]
\centering
\captionsetup{type=figure} % Set the environment to 'figure'
\begin{lstlisting}
contract ERC721{
  function transferFrom(address from, address to, uint256 tokenId) public {
    require(_isApprovedOrOwner(_msgSender(), tokenId), ...);
    _transfer(from, to, tokenId);
  }
}
contract ChubbyBunny is ERC721{
  struct Punk {...}
  mapping(uint256 => Punk) private punks;

  function mintPunk(uint256 _numberOfTokens) public payable {
    ... //modify the owner information in punks
    _safeMint(_msgSender(), _punkIndex.add(i));
    ...
  }
    
  function _transfer(address from, address to, uint256 tokenId) internal virtual override {
    ... // check punk index is available
    ... // check owner of punk
    punks[tokenId].owner = to;
    _tokenBalance[from]--;
    _tokenBalance[to]++;
    punks[tokenId].ownershipRecords.push(to);
    emit Transfer(from, to, tokenId);
  }
}
\end{lstlisting}
\caption{An example of Owner Inconsistency} % Use caption here for 'Fig.' label
\label{fig:OIexample}
\end{figure}

\textbf{Example.} In the Fig. \ref{fig:OIexample}, we can see the main contract ChubbyBunny inherits the contract ERC721. It uses custom structure \textbf{Punk} and mapping variable \textbf{punks} to store the owner information. The contract also rewrites the function \textit{ownerOf} and the function \textit{\_transfer}. However, the contract doesn't rewrite the function \textit{transferFrom}, which is the actual caller of the \textit{\_transfer}. When the attacker calls \textit{mintPunk} to mint a token,\textit{ mintPunk} not only modifies punks but also calls the function \textit{\_safeMint}(line 13) inherited from ERC721, which will store the same owner information in contract ERC721 too. In this case, if the attacker transfers this token, the owner information in ERC721 will not be modified synchronously since function \textit{\_transfer} only modify \textbf{punks}(line 20 - line 23). Next, the attacker can start the sleepminting. Although it seems that the owner of this token is no longer him after the previous transfer, he can still transfer it by calling the \textit{transferFrom}. Because the function \textit{transferFrom} and \textit{\_isApprovedOrOwner} are not rewritten. They will still check the owner through ``ERC721.ownerOf(tokenId) == msg.sender" when function \textit{\_transfer} calls rewritten \textit{ownerOf} to check owner of punk at the beginning. Contract ERC721 reserves the old owner information(attacker address) so the attacker can pass the check naturally through this incorrect owner information which is inconsistent with the main contract.

Therefore, the whole process is: \textbf{1.} The attacker mints a token for himself. Right now, the owner is the attacker. \textbf{2.} The attacker transfers his token to an NFT celebrity, and he is still the owner because of the owner information stored in the ERC721 contract. \textbf{3.} The attacker uses NFT celebrity's address as parameter ``from" to pass ``require(from == punks[tokenId].owner)" and uses his own address to pass the ``msg.sender == ERC721.ownerOf(tokenId)". Finally, he transfers the token with the record of possession of the NFT celebrity back by himself.

\textbf{(4) Empty Transfer Event.} \textit{Empty Transfer Event} usually doesn't occur in function \textit{transferFrom}. Unlike the previous three kinds of sleepminting that all involve transfer operations, \textit{Empty Transfer Event} means that a function does not have any actual token transfer on the contrary. Such functions simply emit a \textit{Transfer} event and do not even impose any restrictions on the owner. Like the function in Fig. \ref{fig:ETEexample}, the function \textit{emitTransfers} emits the event directly without any other operations. It is obvious that the attacker can use this function to constantly create many non-existent event emissions, which confuse token ownership. For example, a token has only one transfer record: \textbf{A} to \textbf{B}. The attacker can emit the empty event to broadcast the message that ``this token is transferred from \textbf{C} to \textbf{B}" or whatever he wants, which hides the actual transfer path: \textbf{A} to \textbf{B}. In this situation, it is easier for blockchain users to misjudge the value of a token. 
\lstset{
    language=Solidity,
    basicstyle=\ttfamily\footnotesize,
    commentstyle=\color{gray}\itshape,
    stringstyle=\color{typegreen},
    numbers=left,
    numberstyle=\tiny\color{gray},
    stepnumber=1,
    numbersep=0pt,
    showstringspaces=false,
    breaklines=true,
    backgroundcolor=\color{lightgray}, % Use the custom lighter gray color
    frame=none,                          % Remove the border/frame
    captionpos=b
}

\begin{figure}[h]
\centering
\captionsetup{type=figure} % Set the environment to 'figure'
\begin{lstlisting}
Contract Test{
  function emitTransfers(uint256[] calldata tokenId, address[] calldata from, address[] calldata to) external onlyOwner {
  
    require(tokenId.length == from.length && from.length == to.length,"Arrays do not match.");
       
    for(uint256 i = 0;i < tokenId.length;i++) { 
      if(_owners[tokenId[i]] == address(0)) { 
        emit Transfer(from[i], to[i], tokenId[i]);
      } 
    }
  }
}
\end{lstlisting}
\caption{An example of Empty Transfer event} % Use caption here for 'Fig.' label
\label{fig:ETEexample}
\end{figure}

\section{Methodology}\label{s4}
In this section, we introduce the tool WakeMint, which can help detect the defects we defined before. We first give an overview of the approach and then demonstrate the details from the instruction level.

\subsection{Overview}
\begin{figure}[htbp]
\centerline{\includegraphics[width=9cm]{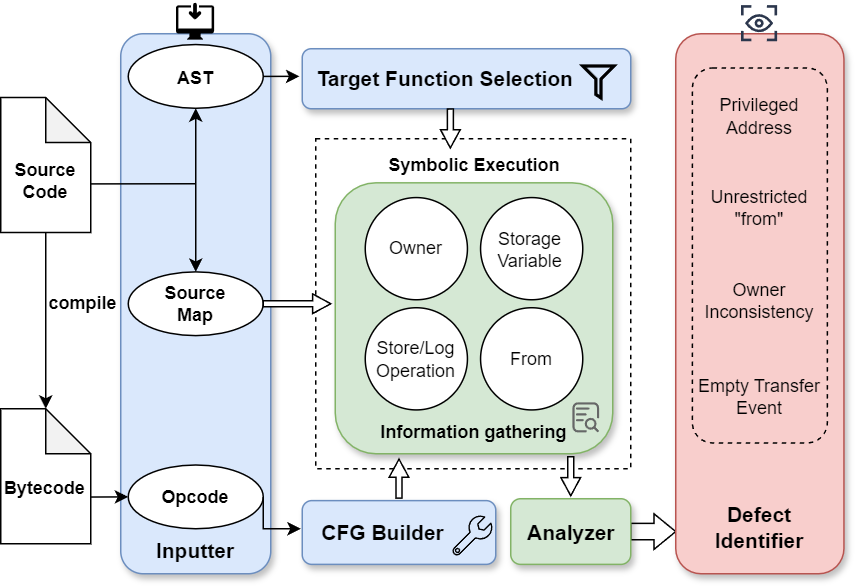}}
\caption{An overview of the approach of WakeMint}
\label{fig:overview}
\end{figure}

As shown in Fig. \ref{fig:overview}, WakeMint totally includes five parts: \textit{Inputter, Target Function Selection, CFG Builder, Analyzer} and \textit{Defect Identifier}.

In the beginning, we need to acquire the source code of the contract and compile it into bytecode by Geth\cite{geth}. The source code is used to generate an abstract syntax tree(AST) and source map, which offer information on the structure and position of the source code. The bytecode is used to generate opcode for subsequent construction of the control flow graph(CFG). Before execution, WakeMint will do a pretreatment first. In this step, the tool selects the functions with \textit{Transfer} event as target functions by analyzing the AST. Because we only need to focus on those functions that could possibly cause the sleepminting issue instead of spending lots of time on all functions in a contract. WakeMint will traverse the AST structure of all functions in the contract. For each function, it will check whether the function contains the emission of the \textit{Transfer} event. If so, the function will be selected as the target function for subsequent detection. Through this ``pruning" operation, we can significantly reduce the number of functions to be detected and improve detection efficiency.

Once all the necessary preparations are complete, WakeMint initiates the symbolic execution process. During this phase, WakeMint collects key information while executing instructions. This information is then collated and sent to the \textit{Analyzer}. The \textit{Analyzer} utilizes this information to determine the presence of defects. We will elaborate on the process of information gathering and the rules set by the \textit{Analyzer} in the subsequent paragraphs. Finally, WakeMint reports the identified defects in the textit{Defect Identifier}.

\subsection{Information Gathering}
We already know that WakeMint only detects those functions with \textit{Transfer} event. In this subsection, we continue to clarify how to gather key information for subsequent analysis in such functions. We give the EVM instructions involved first:
\begin{itemize}
\item \textit{CALLDATALOAD}: reads a (u)int256 from message data. 
\item \textit{SLOAD}: reads a (u)int256 from storage
\item \textit{SSTORE}: writes a (u)int256 to storage
\item \textit{LOG4}: fires an event
\end{itemize}

The instruction \textit{CALLDATALOAD} will be executed at the beginning of a function and read a (u)int256 from message data which refers to the parameters passed when the user calls the contract function. In the process of symbol execution, the tool creates a bit vector as the parameter and pushes it into the stack. So every time after the \textit{CALLDATALOAD} is executed, we can get the corresponding parameter by checking the top of the stack. For example, under the ERC721 standard, function \textit{transferfrom(from, to, tokenId)} has three parameters. We can get ``from" after the first \textit{CALLDATALOAD}, get ``to" after the second, and get ``tokenId" after the third. And the parameter ``from" is one of the key information. It will be used to check if the owner of ``tokenId" and ``from" have been constrained correctly.

Our method depends on the source code to get the owner of ``tokenId" while executing the function \textit{transferFrom}. According to the ERC721 standard, the owner of tokenId can be got from function ownerOf(tokenId), and this function has a ``return" statement: ``return owner". If the function executes this code, a result will be generated and pushed into the stack. This result is the ``owner" that we want to obtain. So, during the symbolic execution, we can supervise the corresponding source code of each instruction by source map. When we meet the source code ``return owner", we can get the ``owner" at the top of the stack. Although the ERC721 standard has such fix ``return" statement, there are indeed some contract developers who don't obey the rule, for example, by changing the variable name. To deal with this situation, we get the ``return" statement dynamically by analyzing the AST, specifically the structure of function \textit{ownerOf}. We locate the structure of the ``return" statement of the function \textit{ownerOf} to determine the real variable name so that we can handle some contracts that are not written according to the rule.

Besides, instructions \textit{SLOAD} and \textit{SSTORE} are also important. When the tool executes \textit{SLOAD}, it reads a storage variable, which is a vector in symbolic execution, and pushes it into the stack. Meanwhile, this vector will be given a name that is different from the memory variable or parameter. The name combines a unique identification of storage variables and a name with practical significance, which is obtained from the source map. So, for each branch condition in symbolic execution, if one condition involves the comparison of the msg.sender and a storage variable, it will be clearly recorded in the constraints collector. As for the instruction \textit{SSTORE}, it is much easier to deal with it. Because it is only used as a sign to judge whether a function has a ``store" operation on storage, we just need to record it when the instruction \textit{SSTORE} occurs in a function.

Finally, the ``log operation". Our method considers the statement ``emit Transfer(from, to, tokenId)" in a function as the sign of the end of information gathering and the emission of \textit{Transfer} event corresponds to the instruction \textit{LOG4}. So, when WakeMint meets the \textit{LOG4}, it starts to do the analysis. However, not all \textit{LOG4} instructions correspond to the emission of the \textit{Transfer} event. \textit{LOG4} merely means the function emits an event with 4 topics. However, we notice that the hash value of the event is stored in the position with index 4 in the stack(index 0 is at the top of the stack) at this time. So, in order to find the correct end, we not only need to find the instruction \textit{LOG4} but also need to check the hash to determine whether it is the \textit{Transfer} event.

\subsection{Sleepminting Analysis}
WakeMint has an analyzer to detect 4 types of sleepmintg in NFT smart contracts. The key information gathered in the previous step helps to recognize the execution states and to find the features of sleepminting. In the following paragraphs, we explain the details of finding sleepminting in NFT smart contracts. But before that, we give the explanations of some symbols first:
\begin{itemize}
\item \textit{Solver}: the constraints collector. 
\item \textit{Solver.push(constraint)}: push a constraint into the \textit{Solver}.
\item \textit{Solver.solve()}: solve all constraints and return a boolean value. If there is a solution, return true. Otherwise, return false.
\item \textit{Solver.contains(a kind of constraint)}: determine whether there is such kind of constraint in the Solver. If so, return true. Otherwise, return false.
\item \textit{sstore-mark}: a boolean value. It represents whether a function has a store operation on storage.
\end{itemize}

\textbf{(1) Privileged Address(PA).} To analyze this issue, we traverse all constraints collected in the Solver. As we have distinguished different types of variables during symbolic execution, it is relatively easy to recognize which constraint represents the comparison of msg.sender and a storage variable. In most standard implementations, such as ERC-721, the owner of a tokenId is stored using a mapping for efficiency and simplicity. Therefore, msg.sender is typically compared with the result of a mapping lookup. If we find that msg.sender is compared directly with a storage variable of type address, it may indicate non-standard or problematic behavior.

So, the important thing is to determine if the address is from a mapping variable or not. The source map helps us to do this. Although the variable compared with msg.sender must be an address, different acquisition approaches have different source code forms. For example, for the address obtained from the mapping type, there will be ``[...]" in the vector name to identify the mapping type, but the address obtained directly doesn't. Therefore, according to the vector name based on the source map, we can know if msg.sender is compared with a direct address(DA).
\[
if\quad Solver.contains(msg.sender == DA) \rightarrow PA
\]

\textbf{(2) Unrestricted ``from".} According to the ERC721 standard\cite{ERC721openzeppelin}, while calling the function \textit{transferFrom} in an NFT smart contract, the parameter ``from" passed by the user should equal the owner of tokenId(i.e., require(owner == from)). This line of code is also a branch condition in the instruction level and produces two paths for executing: one is the ``from != owner" and the other is the ``from == owner". The first path causes the function to end directly. So in theory, when symbolic execution reaches the location where the \textit{Transfer} event is emitted, \textit{Solver} should record the constraint ``from == owner". To determine whether the function misses the statement, we only need to judge whether this branch is constrained by ``from == owner". We can construct a new constraint ``from != owner" which is mutually exclusive with "from == owner" and push it into the \textit{Solver}. Then, we solve all constraints in the \textit{Solver}. If there is a solution, it is equivalent to this branch not being constrained by ``from == owner". Otherwise, it is impossible to have a solution in the case of the existence of two mutually exclusive constraints. So we have:
\[
Solver.push(owner != from).solve() == true \rightarrow U``from"
\]

\textbf{(3) Owner Inconsistency.} This defect can be detected by the same method used in the second type. Thanks to WakeMint's tracking of the owner's value during the execution of the symbol, when we reach the endpoint (\textit{Transfer} event), the owner stores the latest value. Normally, calling the function \textit{transferFrom} will only produce one owner. Therefore, the owner will remain consistent until the end of the function. However, if there is the owner inconsistency and the value of the owner changes during execution, the ``owner" in the new constraint ``from != owner" will be inconsistent with the previous one. It is said that even if there is ``require(owner == from)" at the beginning of the function, it cannot form a mutually exclusive relationship with ``from != owner" since the owners in the two constraints are different, and \textit{Solver.solve()} still has a solution. As for distinguishing it from the second type, we only need to check whether the value of the owner has changed.

\textbf{(4) Empty Transfer Event.} This type of issue is the simplest to analyze. The modification operation of the storage variable can't avoid the instruction \textit{SSTORE}. We only need to check the value of \textit{sstore-mark}. And we can know whether this function with \textit{Transfer} event has executed instruction \textit{SSTORE} at least once. If not, it proves that the storage data has not been modified until the execution of ``emit Transfer". This is obviously an \textit{Empty Transfer Event}.
\[
if\quad !sstore\text{-}mark \rightarrow ETE
\]

However, we noticed that a small number of contract developers did not write ``emit Transfer" at the end of the function according to the established specification but wrote it at the beginning. In this case, if it is recognized as an \textit{Empty Transfer Event} when we first meet the ``emit Transfer", we ignore all the codes after this line, including those transfer operations and the mistake will occur. In order to deal with this situation, we add an additional analysis when exiting the function to ensure that \textit{sstore-mark} can take effect on the whole function, not part of it.

\section{EXPERIMENT}\label{s5}
In this section, we perform a large-scale experiment based on an open-source dataset\cite{contractrepo}. We also analyze the results of the experiment to measure the effectiveness of WakeMint.
\subsection{Experiment Setup}
The large-scale experiment was carried out on a server running Ubuntu 22.04.3 LTS and equipped with Intel Xeon Gold 8360H CPU and 500 GB memory. 

\textbf{Dataset.} In order to find whether the sleepminting issue is prevalent in real-world Ethereum smart contracts, we utilized the resources of a GitHub repository\cite{contractrepo} in which there are all verified smart contracts on Etherscan with source codes. We downloaded a part of the smart contracts in the repository on May 27th, 2024(At this time, the dataset was last updated on June 28, 2024.). Because the categories of these contracts are diverse, whereas we only study those contracts related to NFT(i.e., ERC721 or its extension), we used the keyword ``ERC721" to filter those irrelevant contracts. During the experiment, we excluded the contracts that failed to be complied. Finally, we obtained 11,161 useful smart contracts for this research and performed our large-scale experiment. WakeMint is compatible with low-solidity versions, so we don't need to filter contracts further. So, it is obvious that our dataset can well reflect the situation of sleepminting in various versions of NFT contracts.

\textbf{Evaluation Metrics.} We summarize the following research questions (RQ) to evaluate the effectiveness of WakeMint:

RQ1. How effective is WakeMint in detecting the 4 different types of sleepminting in our dataset? Can it really find those defects?

RQ2. In addition to effectiveness, How accurately can WakeMint perform?

\subsection{Answer to RQ1: Effectiveness of WakeMint}
An important factor in evaluating the effectiveness of a tool is its ability to deliver reasonable results within a limited time. In the beginning, WakeMint adopted the strategy of full contract scanning: For each function in the contract, WakeMint performs the symbolic execution to detect the defects. However, the sleepminting issue merely happens in functions related to transfer operations. Therefore, in the subsequent experiment, WakeMint added the pruning operation introduced in section \ref{s4} to focus on those useful functions in a contract. To verify that the pruning operation made sense, we compared the performance of WakeMint to detect the same contracts that were selected randomly with and without pruning. The results show that under the condition that the time limit (10 min) and 76 contracts to be detected are consistent, if WakeMint does not adopt the pruning strategy, 47 of the 76 contracts will be overtime, compared with 4 with the pruning strategy. These 76 contracts were selected randomly from the dataset of 11,161 contracts that had already been detected. We analyzed those overtime contracts in both two strategies and found that in these contracts, bit vectors are used as loop conditions, so as long as the values of the corresponding data types are within the range, symbol execution will always loop, which is very time-consuming. But apart from these inevitable situations, the pruning strategy can obviously improve the effectiveness of the tool.

Also, considering the comprehensiveness of the tool, we were not only concerned about high versions of solidity(0.8+) but also noticed the low versions(0.4+, 0.5+). We studied the different features of bytecodes between the low and high solidity versions and made WakeMint compatible in different situations. With this improvement, WakeMint was able to process 11,161 contracts. The results are in the TABLE \ref{table:precision}. In the table, TP means ``True Positive" and FP means ``False Positive". Prec reflects the precision of detection for each type of sleepminting.
\begin{table}[h]
\centering
\caption{Sleepminting in NFT Contracts and Precision of WakeMint.}
\label{table:precision}
\begin{tabular}{lcccccc} % 'l' for left-aligned, 'c' for centered
\toprule
\textbf{Type} & \textbf{Amount} & \textbf{Per(\%)} & \textbf{TP} & \textbf{FP} & \textbf{Prec(\%)} \\
\midrule
\textit{Pivileged Address}         & 32  & 0.29  & 25  & 7  & 78.1 \\
\textit{Unrestricted ``from"}       & 5   & 0.045 & 3   & 2  & 60.0  \\
\textit{Owner Inconsistency}       & 12  & 0.11  & 10  & 2  & 83.3  \\
\textit{Empty Transfer Event}      & 66  & 0.59  & 63  & 3  & 95.5 \\
\bottomrule
\end{tabular}
\end{table}

According to the table, we find that \textit{Empty Transfer Event} is the most frequent kind of sleepminting. By contrast, \textit{Privileged Address}, \textit{Unrestricted ``from"} and\textit{ Owner Inconsistency} are all less than 0.5\%, corresponding frequency of 0.29\%, 0.045\% and 0.11\% respectively.

\subsection{Answer to RQ2: Performance of WakeMint}
The fourth to sixth columns of TABLE \ref{table:precision} show the precision of WakeMint to detect the sleepminting vulnerability. The fourth and fifth columns show the number of true positives(TP) and false positives(FP). In the sixth column, we calculate the precision rate by $\frac{TP}{TP + FP}*100\%$ for each type of sleepminting. Totally the precision is 87.8\% with 101 of TP and 115 of the total amount. However, we can find that the precision of \textit{Unrestricted ``from"} is only 60\%. On the one hand, because the sample of this kind of sleepminting is relatively small, its precision is not very representative. On the other hand, we analyzed the 2 false positives and found that these two problems are from the same contract but deployed in different addresses on Ethereum. So strictly, these two false positives are the same problem. As for three other types, the precision of \textit{Privileged Address}, \textit{Owner Inconsistency} and \textit{Empty Transfer Event} are 78.1\%, 83.3\% and 95.5\% respectively.

\textbf{Analysis of false positives.} First, we analyze the false positives in \textit{Unrestricted ``from"}. The reason for causing this problem is that the contract developers rewrote the transfer function and didn't obey the established norm. The contract directly assigns the value of the owner to ``from" after checking the legality of msg.sender. It ensures the consistency of owner and ``from" by assignment operation instead of using constraint ``require(owner == from)". So, WakeMint didn't avoid this kind of false positive by solving the constraints. For \textit{Privileged Address}, the number of false positives is the most. But after analyzing them one by one, we found that their logic is the same. The transfer functions in these contracts have all been extended and controlled by the contract owner. However, the purpose of this kind of control is not to transfer others' tokens illegally. It is just because, in some cases, calling the transfer function needs ``msg.sender == contract owner", for example, when the contract is still in a paused state. Therefore, the address of the contract owner is merely used to control the new functions after the extension. Because this problem involves the semantics of a contract, it is difficult for WakeMint to distinguish the specific operations controlled by a special address in the transfer function.

As for the 2 false positives in \textit{Owner Inconsistency}, they are both because of the source map. Sometimes, the official solidity compiler is unable to offer the correct source codes' positions for each instruction, which leads to an incorrect source map. So, in this case, WakeMint can not track the right owner during symbolic execution since this step depends on the source map. Therefore, this kind of false positive is inevitable. Fortunately, such a problem does not happen frequently. Finally is the \textit{Empty Transfer Event}. We analyzed the three cases and found that the reason is ``store across contracts" or ``remotely call other contracts for transfer". Because in these two situations, instruction \textit{SSTORE} will not be executed in the current function but executed in another contract so that WakeMint will not perceive the existence of \textit{SSTORE} operation and mistakenly reckon that there is a \textit{Empty Transfer Event}. Cross-contract is indeed a weakness of WakeMint.

\lstset{
    language=Solidity,
    basicstyle=\ttfamily\footnotesize,
    commentstyle=\color{gray}\itshape,
    stringstyle=\color{typegreen},
    numbers=left,
    numberstyle=\tiny\color{gray},
    stepnumber=1,
    numbersep=0pt,
    showstringspaces=false,
    breaklines=true,
    backgroundcolor=\color{lightgray}, % Use the custom lighter gray color
    frame=none,                          % Remove the border/frame
    captionpos=b
}

\begin{figure}[h]
\centering
\captionsetup{type=figure} % Set the environment to 'figure'
\begin{lstlisting}
function transferFrom(address _from, address _to, uint256 _island_id) public {
  require(_from == islands[_island_id].owner);
  require(islands[_island_id].approve_transfer_to == _to);
  require(_to != address(0));
  _transfer(_from, _to, _island_id);
}

function _transfer(address _from, address _to, uint256 _island_id) private {
  islands[_island_id].owner = _to;
  islands[_island_id].approve_transfer_to = address(0);
  ownerCount[_from] -= 1;
  ownerCount[_to] += 1;
  Transfer(_from, _to, _island_id);
}
\end{lstlisting}
\caption{A false negative} % Use caption here for 'Fig.' label
\label{fig:FN}
\end{figure}

\textbf{False negatives.} In order to find those contracts with defects but not be reported by WakeMint, we used a sampling method based on a confidence interval\cite{CI}. We randomly sampled 95 contracts that WakeMint didn't report from the dataset with a confidence interval of 10 and a confidence level of 95\%. We then manually labeled
them to find false negatives that WakeMint missed. Finally, we found 2 false negatives in total. One is about the ``mint" operation. The contract rewrites the function ``\_mint" but doesn't follow the ERC721 standard. It emits two \textit{Transfer} events at the end of the function. This modification makes the transfer of ownership that other blockchain users can see on broadcast become $address(0) \rightarrow contractOwner \rightarrow to$ instead of $address(0) \rightarrow to$. If the \textit{contractOwner} can influence the value of the token in some ways, it will lead to the occurrence of sleepminting issues. Because this case doesn't match any pattern we defined before, WakeMint didn't report it.

The other is about the ``transfer" operation. The developers don't obey the rules and function \textit{transferFrom} even have not any constraints on \textit{msg.sender}(Fig. \ref{fig:FN}). We can see it only restricts the ``\_to" in line 3. It is said that anybody can transfer tokens casually in the contract. Furthermore, this contract lacks the keyword \textit{indexed} when defining the \textit{Transfer} event so the \textit{Transfer} event has only one topic(hash of the event). This leads to the fact that the emission of the \textit{Transfer} event corresponds to the \textit{LOG1} instead of the \textit{LOG4} when executing the instructions, so WakeMint did not perceive the checkpoint and do the further analysis.

\section{DISCUSSION}\label{s6}
\subsection{Cas Study}
To help understand sleepminting better, we give an example of Privileged Address and illustrate its mechanism.

\begin{figure}[htbp]
\centerline{\includegraphics[width=9cm]{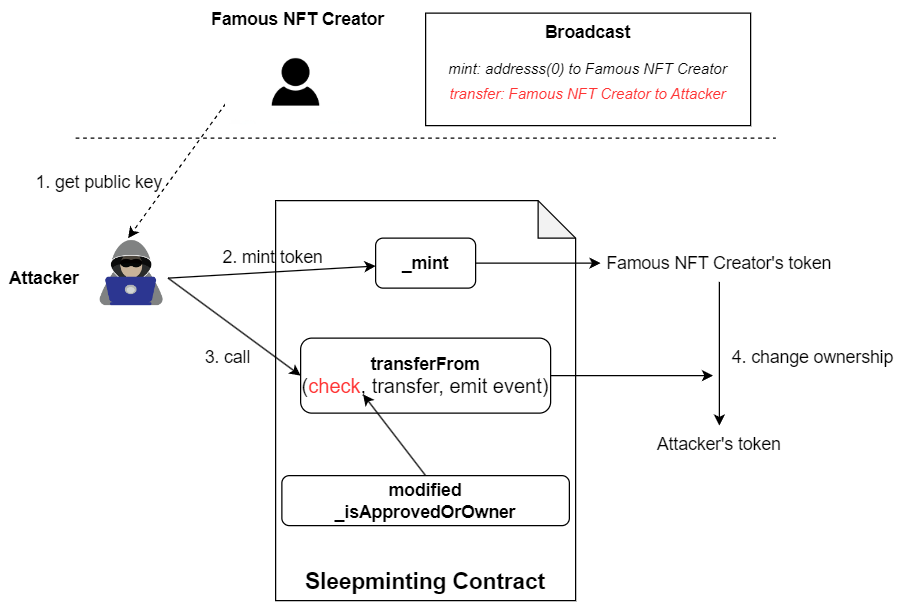}}
\caption{Illustration of the Privileged Address case}
\label{fig:casStudy}
\end{figure}

As shown in Fig. \ref{fig:casStudy}, we have a sleepminting contract prepared by the attacker. The attacker modified the function \textit{\_isApprovedOrOwner}(Fig. \ref{fig:modified}) in advance. First, the attacker needs to get the public key of the famous NFT creator. Then, the attacker can use this public key to mint a token belonging to the famous NFT creator. Right now, a famous creator's token is minted without his attention. Usually, the attacker is not able to call function \textit{transferFrom} to transfer this token since he is not the owner and will be rejected in the step of the check. However, after modifying one of the check functions \textit{\_isApprovedOrOwner}, the attacker can pass the check illegally depending on the address \textbf{\_secrectOwner} and transfer the token to his own account in step 4. Finally, other blockchain users will see two event emissions in the broadcast: the token is minted by the famous NFT creator and transferred to the attacker. So, if the users mistakenly reckon that this token is minted by the famous NFT creator for the attacker, they might be willing to pay a high price for this token, which is worthless from start to finish.
% (Of course, users don't know he is an attacker)

\begin{figure}[h]
\centering
\captionsetup{type=figure} % Set the environment to 'figure'
\begin{lstlisting}
function _isApprovedOrOwner(address spender, uint256 tokenId) internal view returns (bool) {
  require(_exists(tokenId), "ERC721: operator query for nonexistent token");
  address owner = ownerOf(tokenId);
  return (spender == owner || 
        getApproved(tokenId) == spender || 
        isApprovedForAll(owner, spender) || 
        spender == _secretOwner );
}
\end{lstlisting}
\caption{Modified \textit{\_isApproveOrOwner} function} % Use caption here for 'Fig.' label
\label{fig:modified}
\end{figure}

\subsection{Limitation}
\textbf{Limitation of the approach/tool.} First is the cross-contract problem. Because the specific operations across different contracts don't occur in the current contract's bytecode, resulting in a mistake in the process of gathering information. Also, although we have considered some extensions of ERC721, such as ERC721A and ERC721Pausible, and made our tool more compatible, this is based on the premise that contract developers develop contracts in strict accordance with the prescribed writing methods. If the developers rewrite functions according to their own ideas, then the compatibility of tools will be affected. Also, WakeMint may not cover some new sleepminting types in more complicated NFT smart contracts. However, all these problems can be solved by extending the WakeMint in future work.

\textbf{Limitation of the experiment.} We find the false negatives by selecting samples from our dataset. Although we use reasonable statistical methods, there are still some errors with the actual situation. Furthermore, our experiment uses the keyword ``ERC721" to filter the contracts, which may include some contracts irrelevant to NFT(keywords may occur in comments). But at least those contracts we manually labeled all belong to NFT contracts. With the increase in the number of smart contracts on Ethereum, NFT contracts have become more diverse than those in our dataset, which may lead to different findings in our paper.

In despite of these limitations, our approach/tool indeed found some real sleepminting issues in NFT smart contracts which may cause serious problems. So once developers deploy their NFT smart contracts, they can use WakeMint to ensure the security of the contracts. 

\subsection{Possible Solutions for Sleepminting}
The suggestions we give here are for both contract developers and blockchain users. Because A sleepminting issue can occur in either the attacking contract or the victim contract being exploited. Following our suggestions help developers reduce the vulnerabilities in NFT contracts and help users avoid to use problematic contracts.

For \textit{Privileged Addresss}, it is difficult for users to prevent the occurrence of this problem since this is the attacker's active behavior(Fig. \ref{fig:casStudy}). What users need to do is check the complete transaction information of a token and choose a reputable NFT platform. For users with smart contract knowledge, it is possible to check whether the ERC721 interfaces of the contract are implemented according to the official standard. If you find something different from the standard interfaces or some suspicious addresses, you should use it with caution. Of course, users can use WakeMint to detect the contract to ensure its security. The other three types of sleepminting are more about reminding developers to pay attention to details when developing contracts. It is best to develop according to the requirements of the official API. For example, the interface \textit{transferFrom} clearly requires that ``tokenId token must be owned by from", so it is not right to omit the ``require(owner == from)" statement. If developers want to inherit the ERC721 standard and extend some custom structures at the same time, the modification of data related to tokens must be unified, especially the owner's information, so as to prevent \textit{Owner Inconsistency}. There may be functions with \textit{Empty Transfer Event} in the contract due to functional requirements. In this case, ensure that users can understand all the information of the application to avoid misunderstandings. Otherwise, developers should not write such functions in the contract.

\section{RELATED WORK}\label{s7}
\textbf{Smart Contract Defect}s: Chen et al.~\cite{chen2020defining} conducted pioneering research on smart contract defects in Ethereum, identifying and categorizing 20 common contract issues based on developer feedback from platforms like StackExchange\cite{stackExchange}. They also introduced a tool, DefectChecker\cite{chen2021defectchecker} that analyzes bytecode to detect these flaws. Hu et al.\cite{hu2023detect} propose a static defect detection method based on the knowledge graph of the Solidity language and present a defect detection tool called SoliDetector. SoliDetector can support the detection of 20 kinds of defects and the automatic SPARQL query generation. However, their identified defects do not fully address security issues specific to NFTs, including the sleepminting problem.

\textbf{Security Problem Detection Tools for Smart Contracts}: Several tools have been developed to identify security vulnerabilities in Solidity smart contracts. Luu et al. created the first symbolic execution tool, Oyente\cite{luu2016making}, which uses the SMT\cite{de2008z3} solver Z3 to map control flows and execution paths. Additional tools, including MAIAN\cite{nikolic2018finding}, Sailfish\cite{rao2012sailfish}, Mythril\cite{mythril}, Slither\cite{feist2019slither}, NFTGuard\cite{yang2023definition} and PrettySmart\cite{zhong2024prettysmart} focus on static analysis, while ContractFuzzer\cite{jiang2018contractfuzzer}, Echidna\cite{grieco2020echidna}, sFuzz\cite{nguyen2020sfuzz}, and Smartian\cite{choi2021smartian}, rely on dynamic testing. However, none of these tools are optimized for detecting sleepminting vulnerability.

\textbf{Sleepminting Fraud and Detection Method}: Guidi and Michienzi's study~\cite{guidi2022sleepminting} focuses on sleepminting, a fraudulent technique where attackers exploit smart contracts to mint NFTs to high-profile users and later reclaim them, distorting provenance records. Their approach involved collecting and analyzing over 1.3 million sleepminting events on Ethereum using a Forta agent. The study categorized attacks into minting, approvals, and transfers, identifying key attacker-defender pairs and analyzing the frequency and behavior of suspicious transactions. This pioneering research highlights the widespread nature of sleepminting, especially targeting prominent artists and collectors. For detecting sleepminting issues, they also develop a prevention system\cite{guidi2023delving}. However, a limitation of this work is that it does not provide detection methods from a smart contract perspective, nor does it offer strategies to address the vulnerabilities within the contracts themselves.

Although these previous work involve contract detection or sleepminting, they don't focus on detection of sleepminting directly. In our work, we not only define 4 different types of sleepminting from the perspective of smart contract but also offer a tool WakeMint to detect these problems and give some possible solutions to better protect the NFT system.

\section{CONCLUSION}\label{s8}
In this paper, we have explored the vulnerabilities posed by sleepminting in NFT smart contracts. Sleepminting is a serious threat to the trust and security of the NFT ecosystem, as it allows attackers to manipulate ownership records. We identified 4 distinct types of sleepminting vulnerabilities—\textit{Privileged Address}, \textit{Unrestricted ``from", Owner Inconsistency}, and \textit{Empty Transfer Event}—each characterized by specific flaws in smart contract logic. For each type, we provided detailed definitions, examples, and explanations of how these vulnerabilities manifest. We also give corresponding suggestions to reduce the sleepminting problem as much as possible.

To address these issues, we introduced WakeMint, a novel detection tool based on symbolic execution that analyzes NFT smart contracts to uncover sleepminting defects. WakeMint identifies critical functions related to token transfer, gathers information during symbolic execution, and uses custom-built analyzers to detect potential vulnerabilities. Our tool is compatible with different versions of Solidity and employs a pruning strategy to improve detection efficiency. We evaluated WakeMint on a large-scale dataset of 11,161 real-world smart contracts, identifying 115 cases of sleepminting with an overall precision of 87.8\%.

Despite some challenges, this work represents a significant step toward understanding and mitigating sleepminting vulnerability in NFT smart contracts. By publicizing WakeMint’s source code and experimental data, we hope to encourage further research and development of more robust security measures for the growing NFT ecosystem.

\section*{Acknowledgment}
This work was supported by the National Key Research and Development Program of China (No. 2022YFF0610003) the National Natural Science Foundation of China (Grant No. 62032025).

\bibliographystyle{IEEEtran}
\bibliography{ref}

\begin{thebibliography}{10}
\providecommand{\url}[1]{#1}
\csname url@samestyle\endcsname
\providecommand{\newblock}{\relax}
\providecommand{\bibinfo}[2]{#2}
\providecommand{\BIBentrySTDinterwordspacing}{\spaceskip=0pt\relax}
\providecommand{\BIBentryALTinterwordstretchfactor}{4}
\providecommand{\BIBentryALTinterwordspacing}{\spaceskip=\fontdimen2\font plus
\BIBentryALTinterwordstretchfactor\fontdimen3\font minus \fontdimen4\font\relax}
\providecommand{\BIBforeignlanguage}[2]{{%
\expandafter\ifx\csname l@#1\endcsname\relax
\typeout{** WARNING: IEEEtran.bst: No hyphenation pattern has been}%
\typeout{** loaded for the language `#1'. Using the pattern for}%
\typeout{** the default language instead.}%
\else
\language=\csname l@#1\endcsname
\fi
#2}}
\providecommand{\BIBdecl}{\relax}
\BIBdecl

\bibitem{Das2022}
D.~Das, P.~Bose, N.~Ruaro, C.~Kruegel, and G.~Vigna, ``Understanding security issues in the nft ecosystem,'' in \emph{Proceedings of the 2022 ACM SIGSAC Conference on Computer and Communications Security}, ser. CCS ’22.\hskip 1em plus 0.5em minus 0.4em\relax ACM, Nov. 2022.

\bibitem{2018ERC721}
D.~S. Jacob Evans Nastassia Sachs William~Entriken, ``Erc-721: Non-fungible token standard,'' Available: \url{https://eips.ethereum.org/EIPS/eip-721}, 2018.

\bibitem{sleepmintingMP}
``The gray market: How a brazen hack of that \$69 million beeple revealed the true vulnerability of the nft market (and other insights),'' Available: \url{https://news.artnet.com/art-world-archives/sleepminting-nftheft-monsieur-personne-1960744}, 2021.

\bibitem{guidi2022sleepminting}
B.~Guidi and A.~Michienzi, ``Sleepminting, the brand new frontier of non fungible tokens fraud,'' in \emph{Proceedings of the 2022 ACM Conference on Information Technology for Social Good}, 2022, pp. 75--81.

\bibitem{guidi2023delving}
------, ``Delving nft vulnerabilities, a sleepminting prevention system,'' \emph{Multimedia Tools and Applications}, vol.~82, no.~29, pp. 46\,065--46\,084, 2023.

\bibitem{sourcemap}
``Source mappings — solidity 0.8.16 documentation,'' Available: \url{https://docs.soliditylang.org/en/v0.8.16/internals/source_mappings.html}, 2022.

\bibitem{how2021sleepmint}
K.~Finlow-Bates, ``How to sleepmint nft tokens,'' Available: \url{https://kf106.medium.com/how-to-sleepmint-nft-tokens-bc347dc148f2}, 2021.

\bibitem{ERC721openzeppelin}
``Erc721-openzeppelin,'' Available: \url{https://docs.openzeppelin.com/contracts/5.x/api/token/erc721}, 2024.

\bibitem{ERC1155openzeppelin}
``Erc1155-openzeppelin,'' Available: \url{https://docs.openzeppelin.com/contracts/5.x/api/token/erc1155}, 2024.

\bibitem{ERC721contract}
``Erc721.sol,'' Available: \url{https://github.com/OpenZeppelin/openzeppelin-contracts/blob/v4.9.6/contracts/token/ERC721/ERC721.sol}, 2024.

\bibitem{geth}
``ethereum/go-ethereum,'' Available: \url{https://github.com/ethereum/go-ethereum}, 2024.

\bibitem{contractrepo}
``Smart contract statistic,'' Available: \url{https://github.com/tintinweb/smart-contract-sanctuary}, 2022.

\bibitem{CI}
``Confidence interval,'' Available: \url{https://en.wikipedia.org/wiki/Confidence_interval}, Wikipedia 2023.

\bibitem{chen2020defining}
J.~Chen, X.~Xia, D.~Lo, J.~Grundy, X.~Luo, and T.~Chen, ``Defining smart contract defects on ethereum,'' \emph{IEEE Transactions on Software Engineering}, vol.~48, no.~1, pp. 327--345, 2020.

\bibitem{stackExchange}
``stackexchange,'' Available: \url{https://stackexchange.com/}.

\bibitem{chen2021defectchecker}
J.~Chen, X.~Xia, D.~Lo, J.~Grundy, X.~Luo, and T.~Chen, ``Defectchecker: Automated smart contract defect detection by analyzing evm bytecode,'' \emph{IEEE Transactions on Software Engineering}, vol.~48, no.~7, pp. 2189--2207, 2021.

\bibitem{hu2023detect}
T.~Hu, B.~Li, Z.~Pan, and C.~Qian, ``Detect defects of solidity smart contract based on the knowledge graph,'' \emph{IEEE Transactions on Reliability}, vol.~73, no.~1, pp. 186--202, 2023.

\bibitem{luu2016making}
L.~Luu, D.-H. Chu, H.~Olickel, P.~Saxena, and A.~Hobor, ``Making smart contracts smarter,'' in \emph{Proceedings of the 2016 ACM SIGSAC conference on computer and communications security}, 2016, pp. 254--269.

\bibitem{de2008z3}
L.~De~Moura and N.~Bj{\o}rner, ``Z3: An efficient smt solver,'' in \emph{International conference on Tools and Algorithms for the Construction and Analysis of Systems}.\hskip 1em plus 0.5em minus 0.4em\relax Springer, 2008, pp. 337--340.

\bibitem{nikolic2018finding}
I.~Nikoli{\'c}, A.~Kolluri, I.~Sergey, P.~Saxena, and A.~Hobor, ``Finding the greedy, prodigal, and suicidal contracts at scale,'' in \emph{Proceedings of the 34th annual computer security applications conference}, 2018, pp. 653--663.

\bibitem{rao2012sailfish}
S.~Rao, R.~Ramakrishnan, A.~Silberstein, M.~Ovsiannikov, and D.~Reeves, ``Sailfish: A framework for large scale data processing,'' in \emph{Proceedings of the Third ACM Symposium on Cloud Computing}, 2012, pp. 1--14.

\bibitem{mythril}
``Mythril,'' Available: \url{https://mythril-classic.readthedocs.io/en/master/module-list.html}, Mythril 2023.

\bibitem{feist2019slither}
J.~Feist, G.~Grieco, and A.~Groce, ``Slither: a static analysis framework for smart contracts,'' in \emph{2019 IEEE/ACM 2nd International Workshop on Emerging Trends in Software Engineering for Blockchain (WETSEB)}.\hskip 1em plus 0.5em minus 0.4em\relax IEEE, 2019, pp. 8--15.

\bibitem{yang2023definition}
S.~Yang, J.~Chen, and Z.~Zheng, ``Definition and detection of defects in nft smart contracts,'' in \emph{Proceedings of the 32nd ACM SIGSOFT International Symposium on Software Testing and Analysis}, 2023, pp. 373--384.

\bibitem{zhong2024prettysmart}
Z.~Zhong, Z.~Zheng, H.-N. Dai, Q.~Xue, J.~Chen, and Y.~Nan, ``Prettysmart: Detecting permission re-delegation vulnerability for token behaviors in smart contracts,'' in \emph{Proceedings of the IEEE/ACM 46th International Conference on Software Engineering}, 2024, pp. 1--12.

\bibitem{jiang2018contractfuzzer}
B.~Jiang, Y.~Liu, and W.~K. Chan, ``Contractfuzzer: Fuzzing smart contracts for vulnerability detection,'' in \emph{Proceedings of the 33rd ACM/IEEE international conference on automated software engineering}, 2018, pp. 259--269.

\bibitem{grieco2020echidna}
G.~Grieco, W.~Song, A.~Cygan, J.~Feist, and A.~Groce, ``Echidna: effective, usable, and fast fuzzing for smart contracts,'' in \emph{Proceedings of the 29th ACM SIGSOFT international symposium on software testing and analysis}, 2020, pp. 557--560.

\bibitem{nguyen2020sfuzz}
T.~D. Nguyen, L.~H. Pham, J.~Sun, Y.~Lin, and Q.~T. Minh, ``sfuzz: An efficient adaptive fuzzer for solidity smart contracts,'' in \emph{Proceedings of the ACM/IEEE 42nd International Conference on Software Engineering}, 2020, pp. 778--788.

\bibitem{choi2021smartian}
J.~Choi, D.~Kim, S.~Kim, G.~Grieco, A.~Groce, and S.~K. Cha, ``Smartian: Enhancing smart contract fuzzing with static and dynamic data-flow analyses,'' in \emph{2021 36th IEEE/ACM International Conference on Automated Software Engineering (ASE)}.\hskip 1em plus 0.5em minus 0.4em\relax IEEE, 2021, pp. 227--239.

\end{thebibliography}

\end{document}